\documentclass[a4paper,11pt]{article}
\usepackage{pos}
\usepackage{wrapfig}
\usepackage{adjustbox}
\usepackage{physics}
\usepackage{lineno}

\begin{document}

\title{IceCube Search for MeV Neutrinos from Mergers using Gravitational Wave Catalogs}     
\ShortTitle{IceCube Search for MeV Neutrinos from Mergers using GW}

\author{Nora Valtonen-Mattila for the IceCube Collaboration\footnote{\url{https://icecube.wisc.edu/collaboration/authors/}} }

% \author[a]{Nora Valtonen-Mattila}
% \author[b]{Spencer Griswold}
% \author*[b]{Segev BenZvi}

% \affiliation[a]{Department of Physics and Astronomy, Uppsala University\\Box 516, S-75120 Uppsala, Sweden}

% \affiliation[b]{Department of Physics and Astronomy, University of Rochester\\Rochester, NY 14627, USA}

% Uncomment \forColl{coll.name} below to add "for the XXX Collaboration" to the authors list. 
% In this case, you also have to uncomment the lines after "%Full authors list" below and include the full authors list,
%\forColl{coll.name} % W/O "Collaboration"

\emailAdd{nora.valtonen-mattila@icecube.wisc.edu}

\abstract{We report on a search using the IceCube Neutrino Observatory for MeV neutrinos from compact binary mergers detected through gravitational waves during the LIGO-Virgo-KAGRA (LVK) O1, O2, and O3 observing runs. The search focuses on events involving at least one candidate neutron star, such as binary neutron star (BNS) and neutron star–black hole (NSBH) mergers, which may produce a burst of thermal neutrinos due to the hot and dense conditions created during the merger. We looked for short-time increases in IceCube’s detector activity around each gravitational-wave event, using four time windows centered on the merger time. We also performed a binomial test for two populations, those with and without at least one neutron star. No significant excess of neutrinos was found. We set upper limits on the MeV neutrino flux for each event, and we place constraints on MeV neutrino emission from mergers that have at least one neutron star. We showcase upper limits for GW170817, the first confirmed BNS merger, providing one of the strongest limits to date on MeV neutrino emission from such sources.

\vspace{4mm}
{\bfseries Corresponding author:}
Nora Valtonen-Mattila$^{1}*$\\
{$^{1}$ \itshape 
Fakultät für Physik \& Astronomie, Ruhr-Universität Bochum, D-44780 Bochum, Germany
}\\
$^*$ Speaker

\ConferenceLogo{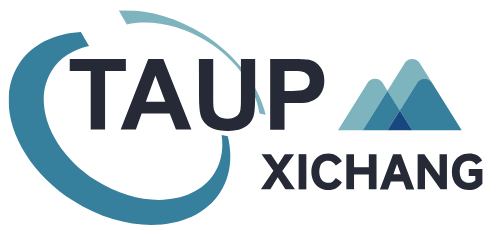}

XVIII International Conference on Topics in Astroparticle and Underground Physics 2023

\FullConference{%
XIX International Conference on Topics in Astroparticle and Underground Physics 2025 (TAUP 2025)\\
  24 August -- 30 August, 2025\\
  Xichang, China}
}

%% \tableofcontents
\maketitle

\section{Introduction}

The IceCube Neutrino Observatory~\cite{IceCube:2016zyt} is a detector located at the South Pole that instruments a cubic kilometer of Antarctic ice to observe neutrinos through the Cherenkov light emitted by energetic charged particles generated in nearby neutrino interactions. The array comprises 5160 digital optical modules (DOMs) mounted on 86 vertical cables (strings) deployed deep within the ice, with an average horizontal spacing of about 125 m. Although IceCube was primarily designed to detect high-energy neutrinos above the GeV scale, it also has sensitivity to intense bursts of low-energy neutrinos in the MeV range~\cite{IceCube:2011cwc}.

\smallskip

Transient astrophysical phenomena, including core-collapse supernovae (CCSNe) and certain binary systems, are predicted to emit quasi-thermal neutrinos with average energies $\expval{E_\nu(t)} \approx 10$--$30~\mathrm{MeV}$ ~\cite{Horiuchi:2018ofe, Merger:2020abc}. At these energies, neutrino interactions in ice occur mainly through inverse beta decay (IBD), $\bar{\nu_e} + p \rightarrow e^+ + n$~\cite{IceCube:2011cwc}, with smaller contributions (a few percent) from alternative processes such as elastic scattering. The resulting positron produces Cherenkov photons along its short track, with a mean length of $0.56~\mathrm{cm}\times(E_{e^+}/\mathrm{MeV})$. Given the relatively large separation between optical modules and the absorption length of light in ice, many of these photons do not reach a DOM. Consequently, the resulting signal is an ensemble of single, uncorrelated hits lacking directional information. Each DOM exhibits an average noise rate of approximately 540~Hz~\cite{IceCube:2011cwc}, dominated by radioactive decays within the glass pressure housing, with a smaller, seasonally varying contribution from atmospheric muons~\cite{IceCube:2011cwc}. While these individual hits cannot be distinguished from noise, low-energy neutrino detection relies on the collective rate enhancement observed across the array relative to the expected background level at a given time (see Fig.~1).

\begin{wrapfigure}{r}{0.55\textwidth}
    \includegraphics[width=\linewidth]{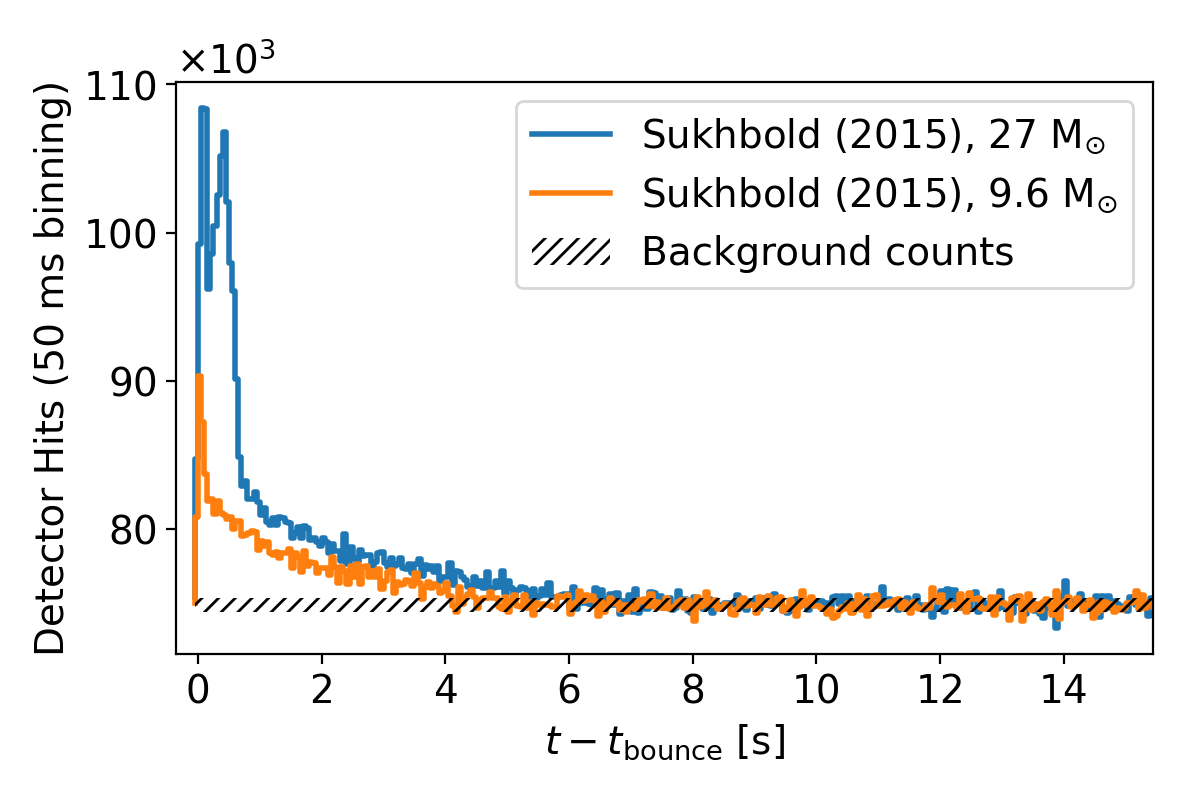}
    \caption[font=large]{Simulation of detector counts for a Galactic CCSN at a distance of 10 kpc, binned in 50 ms, assuming two different masses, as a function of the time from core bounce. From Ref.~\cite{IceCube:2023MeVNora}}
    \vspace{-9pt}
    \label{fig:rate increase}
\end{wrapfigure}

\section{Thermal Neutrinos from Transients}\label{sec:Thermal neutrinos}

Astrophysical transients emit a variety of observables, offering complementary insights into the physical conditions and dynamics of their source environments. Among these, mergers involving neutron stars (NS) represent particularly compelling candidates for joint observations of gravitational waves and thermal (MeV-scale) neutrinos~\cite{Merger:2022abc}. Such events are predicted to be extremely luminous in neutrinos, with electron antineutrino luminosities on the order of $10^{51}$--$10^{53}$~erg/s~\cite{Merger:2020abc}. Binary mergers containing at least one neutron star, such as binary neutron star (BNS) and neutron star–black hole (NSBH) systems, are expected to produce short-lived neutrino bursts characterized by mean energies $\expval{E_\nu(t)} \approx 10$--$30~\mathrm{MeV}$~\cite{Merger:2020abc} and durations ranging from milliseconds to seconds. While CCSNe are also expected to produce thermal neutrinos, no CCSN has occurred sufficiently nearby and coincident with gravitational-wave observations in the current detector era, motivating our focus on mergers. In these proceedings we will describe the search for thermal neutrinos from such systems.

\section{Detection of MeV Neutrino Bursts in IceCube}\label{sec:online_system}

IceCube is sensitive to MeV-scale neutrino bursts through a collective increase in the single photoelectron hit rate across its DOMs. The Supernova Data Acquisition System (SNDAQ) continuously monitors these rates in real time, looking for statistically significant deviations $\Delta_\mu$ in different search bins $r_i$ (signal window) in a sliding time window configuration, relative to the expected background mean $\mu_i$ and standard deviation $\sigma_i$. A test statistic describing the deviation from the expected background is defined as

\begin{equation}\label{eq:delta_mu}
\xi = \frac{\Delta\mu}{\sigma_{\Delta\mu}},\quad \text{where }\ 
\Delta\mu = \sigma_{\Delta\mu}^2 \sum_{i=1}^{N_\mathrm{DOM}}
\frac{\epsilon_i (r_i - \langle r_i \rangle)}{\sigma_i^2},\quad \text{and }\ 
\sigma_{\Delta\mu} = \left( \sum_{i=1}^{N_\mathrm{DOM}} \frac{\epsilon_i^2}{\sigma_i^2} \right)^{-1}.
\end{equation}

\noindent Here, $\Delta\mu$ represents the maximum-likelihood estimator of the collective rate increase across all DOMs, weighted by each DOM’s relative detection efficiency $\epsilon_i$~\cite{IceCube:2011cwc,IceCube:2023sn}. In the absence of correlated noise, the test statistic $\xi$ would follow a Gaussian distribution. However, contributions from cosmic rays, thermal noise, and radioactive decays broaden this distribution. To mitigate such effects, a deadtime of approximately $250~\mu$s is applied during data acquisition, which suppresses short-time-correlated noise and reduces the average DOM rate from 540~Hz to 286~Hz.

\medskip

Residual correlations, particularly those arising from atmospheric muons, are further corrected by applying a linear correction to the test statistic:

\begin{equation}
    \xi_{\mathrm{corr}} = \xi - b \cdot R_\mu - a,
\end{equation}

\noindent where $b$ is the slope, $a$ is the offset, and $R_\mu$ denotes the muon-induced hit rate. The latter varies seasonally, peaking during the austral summer, although its contribution to the total detector rate remains at the percent level~\cite{IceCube:2011cwc}. In addition to real-time monitoring, SNDAQ generates a dedicated dataset (SN Data) containing 0.5~s binned detector rates for offline analyses.

 \begin{figure}
    \centering
    \includegraphics[width=0.8\linewidth]{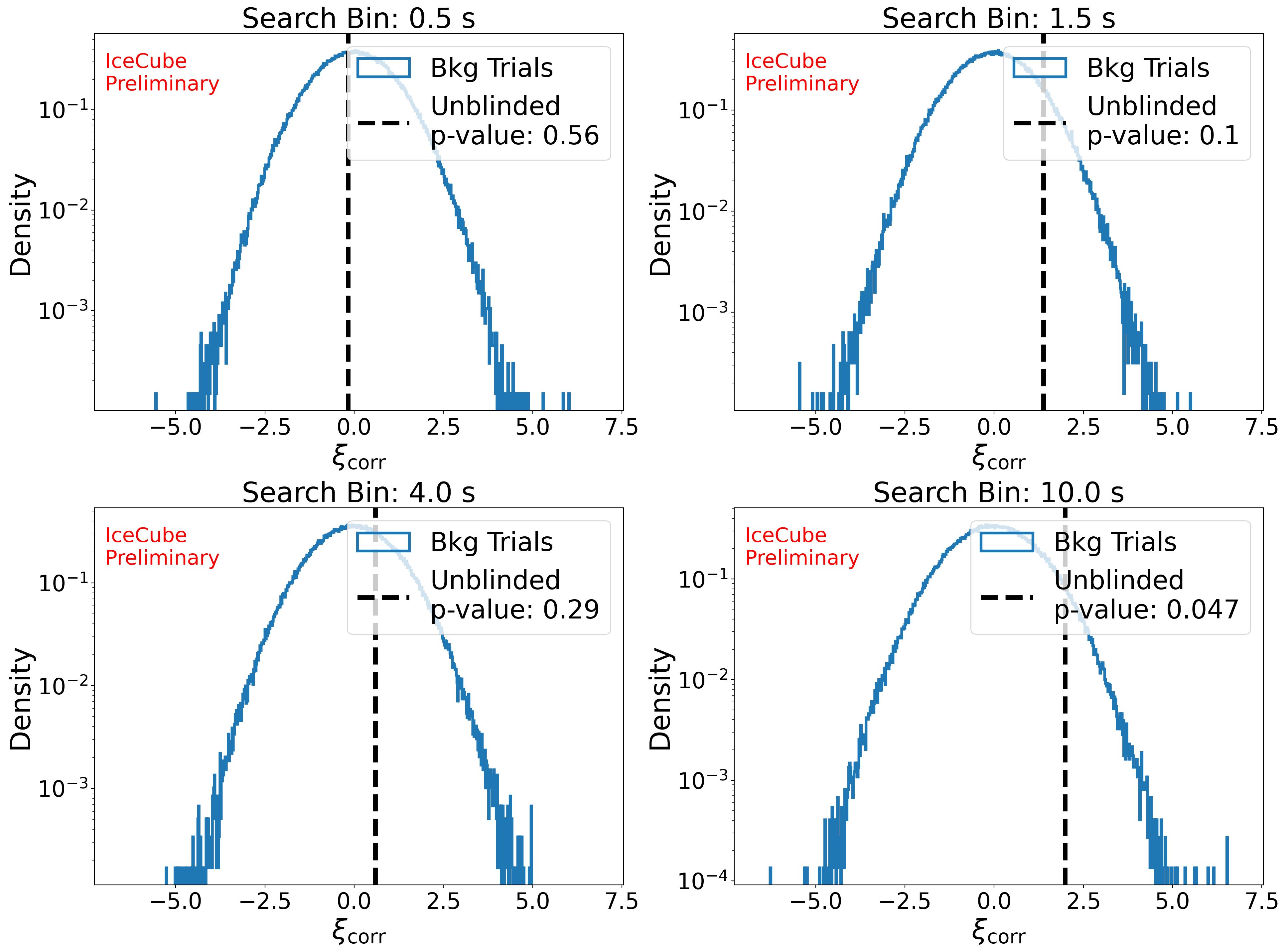}
    \caption[font=large]{TS Distribution for GW 170817 for the four search windows, where the blue distribution is for the muon-corrected TS $\xi_{\mathrm{corr}}$. The black dashed line represent the on-time observed value (unblinded), with the p-value shown in the legend.}
    \vspace{-10pt}
    \label{fig:BKG_Dist_GW170817}
\end{figure}

\section{Searching for Thermal Neutrinos from GW Alerts}

To search for MeV-scale neutrinos temporally correlated with gravitational-wave (GW) events, we used the LIGO–Virgo–KAGRA (LVK) catalogs from observing runs O1 through O3 \citep{LVK:2019a, LVK:2024a, LVK:2023a} to provide the timing information for the search. These cataloged events serve as reference points to define our search start and windows in IceCube. The analysis uses the SN data dataset described in Sec.~\ref{sec:online_system}, which contains 0.5~s-binned detector rates from all DOMs.

\medskip

We adopt a frequentist "on and off-time" approach to quantify possible rate enhancements coincident with GW triggers. For each GW event, the \emph{on-time} window is defined as the time interval containing the GW trigger, while the \emph{off-time} region corresponds to periods where no astrophysical signal is expected. The off-time dataset spans $\pm 24$~hours around the trigger time, excluding the on-time region, and serves as the background sample. We then apply the analysis ($\xi_{\mathrm{corr}}$) on those datasets.

\medskip

To account for uncertainties in the duration of the neutrino emission, we perform the search using four distinct time windows centered on the GW trigger: 0.5, 1.5, 4.0, and 10~s. These correspond to burst timescales predicted for various transient models, ranging from short-lived merger-driven emissions as well as time-of-flight delays. For each of these window durations, a background distribution of the test statistic ($\xi_{\mathrm{corr}}$) is obtained from the off-time data, resulting in four background samples per GW event (see Fig.~\ref{fig:BKG_Dist_GW170817}). The significance of any observed excess in the on-time data is then evaluated relative to the corresponding background distribution.

\medskip

To assess potential population-level signals, we apply a binomial test to the set of p-values from GW events. This test evaluates whether the number of significant events exceeds the expectation from background fluctuations across the entire sample. The GW catalog is divided into two subsets: (i) events classified as mergers containing at least one neutron star candidate (i.e., BNS or NSBH systems), and (ii) those classified as binary black hole (BBH) mergers, which are not expected to produce detectable thermal neutrino emission. The binomial test is applied separately to these two populations. The expected distribution of the binomial test statistic is obtained empirically by applying the same procedure to off-time data, thereby enabling a fully frequentist interpretation of the results. 

\medskip

This analysis framework provides us with a model-independent search for temporally correlated MeV neutrino signals associated with GW events and allows for sensitivity estimates across a range of burst durations and source classes.

\section{Results}

\begin{wrapfigure}{r}{0.62\textwidth}
    \vspace{-\baselineskip} % or -10pt
    \includegraphics[width=\linewidth]
    {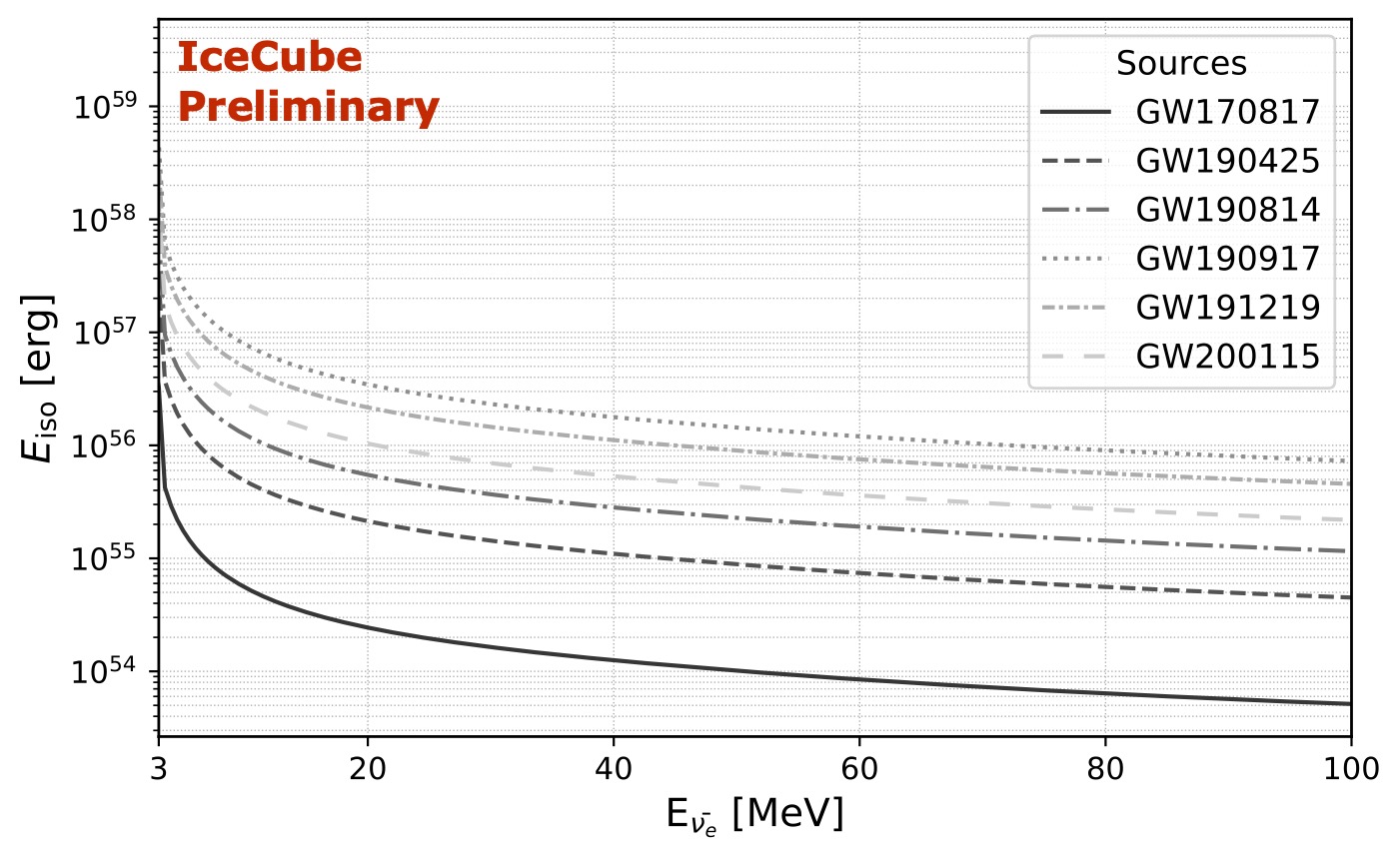}
    \caption[font=large]{Upper limit on the electron antineutrino isotropic equivalent energy at source for mergers involving at least one NS, assuming a monochromatic spectrum with energies between 3 and 100 MeV.}
    \label{fig:UL_Mono}
\end{wrapfigure}
A total of 83 GW alerts were analyzed, consisting of 77 binary black hole (BBH) and 6 neutron-star–containing (BNS/NSBH) mergers. For each individual source, we obtained the p-value for each window and selected the lowest p-value as the result for that source. After applying corrections for the trials factor associated with selecting the most significant window and for the total number of sources examined, all resulting p-values exceed 0.65, indicating no significant excess above background expectations.

\medskip

At the population level, we applied the binomial test separately to the BNS/NSBH and BBH subsets. After accounting for the look-elsewhere effect, we obtain corrected p-values of 0.12 for the BNS/NSBH population and 0.81 for the BBH population. Therefore, no statistically significant correlation between MeV neutrino emission and GW events is observed. Consequently, we set upper limits on the neutrino emission from these sources, with the monochromatic upper limits for the BNS/NSBH population shown in Fig.~\ref{fig:UL_Mono}.

\section{Conclusions}\label{sec:conclusions}

IceCube is capable of detecting bursts of MeV-scale neutrinos through collective rate deviations across its optical modules. We have presented the first IceCube constraints on MeV neutrino emission temporally associated with GW sources. Using data from the LVK O1–O3 observing runs, we searched for thermal neutrino signals coincident with 83 GW alerts across four time windows. No significant excess was observed in either individual events or population-level analyses. Consequently, we set upper limits on the thermal neutrino emission from BNS/NSBH mergers, which represent the most promising GW sources for such signals.

\bibliographystyle{ICRC}
\bibliography{references}

%% Full authors list (ONLY FOR COLLABORATIONS)
\clearpage
%\section*{Full Authors List: \Coll\ Collaboration}
%
%\noindent \textbf{Note comment afterwards:} Collaborations have the possibility to provide an authors list in xml format which will be used while generating the DOI entries making the full authors list searchable in databases like Inspire HEP. For instructions please go to icrc2021.desy.de/proceedings or contact us under icrc2021proc@desy.de.\\
%
%\scriptsize
%\noindent
%first.author$^1$, 
%second.author$^2$, 
%third.author$^3$ % .... more names
%and 
%last.author$^{n}$ \\
%
%\noindent
%$^1$first.affiliation.
%$^2$second.affiliation. % .... more affiliation
%$^{m}$last.affiliation.

%\input{authorlist_icecube.tex}

\end{document}